\documentclass{aastex701}

\usepackage{subcaption}
\usepackage{graphicx}
\graphicspath{{./}}
\begin{document}

\title{A Stable Mineral Fingerprint in the Fading Warm Debris Disk around HD 15407A}

\correspondingauthor{Shu Wang}
\author[orcid=0000-0003-4489-9794]{Shu Wang}
\affiliation{CAS Key Laboratory of Optical Astronomy, National Astronomical Observatories, Chinese Academy of Sciences, Beijing 100101, P.\,R.\,China}
\affiliation{School of Astronomy and Space Science, University of the Chinese Academy of Sciences, Beijing, 100049, P.\,R.\,China}
\email{shuwang@nao.cas.cn}

\author{Zhuochao Huang}
\affiliation{CAS Key Laboratory of Optical Astronomy, National Astronomical Observatories, Chinese Academy of Sciences, Beijing 100101, P.\,R.\,China}
\affiliation{School of Space and Earth Sciences, Beihang University, Beijing 102206, P.\,R.\,China}
\email{hzc2330@buaa.edu.cn}

\author[orcid=0000-0001-7084-0484]{Xiaodian Chen}
\affiliation{CAS Key Laboratory of Optical Astronomy, National Astronomical Observatories, Chinese Academy of Sciences, Beijing 100101, P.\,R.\,China}
\affiliation{School of Astronomy and Space Science, University of the Chinese Academy of Sciences, Beijing, 100049, P.\,R.\,China}
\affiliation{Institute for Frontiers in Astronomy and Astrophysics, Beijing Normal University, Beijing 102206, P.\,R.\,China}
\email{chenxiaodian@nao.cas.cn}

\author{Junming He}
\affiliation{Institute for Frontiers in Astronomy and Astrophysics, Beijing Normal University, Beijing 102206, P.\,R.\,China}
\affiliation{School of Physics and Astronomy, Beijing Normal University, Beijing 100875, China}
\affiliation{CAS Key Laboratory of Optical Astronomy, National Astronomical Observatories, Chinese Academy of Sciences, Beijing 100101, P.\,R.\,China}
\email{202421101128@mail.bnu.edu.cn}

\author{Haomiao Huang}
\affiliation{CAS Key Laboratory of Optical Astronomy, National Astronomical Observatories, Chinese Academy of Sciences, Beijing 100101, P.\,R.\,China}
\affiliation{School of Astronomy and Space Science, University of the Chinese Academy of Sciences, Beijing, 100049, P.\,R.\,China}
\email{huanghm@bao.ac.cn}
\begin{abstract}

Extreme debris disks provide time-domain probes of rocky planet assembly after giant impacts. We compare Spitzer and JWST spectra of the warm debris disk around HD 15407A, obtained 14.95 yr apart. Across 6--25~$\mu$m, the stellar-subtracted mid-infrared excess declined by $14.0\pm1.2\%$, while the mineral fingerprint is nearly unchanged. Within the adopted common reduced mineral basis, Mg-rich pyroxene dominates the normalized fitted mineral weights at about 65\%, while silica and forsterite contribute about 22--23\% and 12--13\%, respectively. In both epochs, about 93--94\% of the fitted mineral weight is carried by 5.0~$\mu$m grains. The fitted optically thin surface-component amplitude drops by 32\%, consistent with reduced mineral-emitting flux from the disk surface rather than the appearance of a new dust composition. A simple collisional-cascade guide fitted to the IRAS-to-JWST multi-epoch ratios gives $t_c=87^{+17}_{-12}$ yr, for which the Spitzer-normalized dust-excess ratio reaches 0.5 after one $t_c$. HD 15407A is consistent with an evolved post-impact reservoir with limited recent supply of very small grains: its rocky mineral fingerprint has remained stable since at least 2008, while its warm mineral-emitting flux is gradually fading over several hundred years.

\end{abstract}

\keywords{\uat{Debris disks}{363} --- \uat{Circumstellar dust}{236} --- \uat{Infrared spectroscopy}{2285} --- \uat{James Webb Space Telescope}{2291} --- \uat{Time domain astronomy}{2109} --- \uat{Planet formation}{1241} --- \uat{F stars}{519}}

\section{Introduction}\label{sec:intro} 

Debris disks are tenuous, usually gas-poor circumstellar environments observed over a wide range of stellar ages, primarily composed of micron- to millimeter-sized secondary dust generated by the continuous collisions of planetesimals \citep{Wyatt2008,Krivov2010,Hughes2018}. Governed by stellar radiation pressure, Poynting-Robertson (P-R) drag, and mutual collisions, many debris disks can be described as quasi-steady collisional cascades. A small subset, however, shows fractional infrared luminosities and warm dust levels far above steady-state expectations. These extreme debris disks (EDDs) are generally interpreted as transient systems associated with recent giant impacts among rocky bodies, often during the late stages of terrestrial planet formation \citep{Balog2009,Meng2014,Lisse2009,Jackson2014,Moor2021}.

Infrared time variability provides a direct way to identify the transient component of EDD dust. Multi-epoch Spitzer and WISE monitoring has shown that some EDDs undergo strong 3--5~$\mu$m changes on monthly to yearly timescales, whereas others show slower decade-scale evolution or comparatively stable short-wavelength emission \citep{Meng2015,Su2019}. HD 166191 illustrates both the diagnostic value and the classification ambiguity of such variability: although it may be closer to a gas-rich transition or hybrid disk than to a classical gas-poor debris disk \citep{Kennedy2014,Worthen2026}, its rapid mid-infrared brightening accompanied by optical dimming was interpreted as a projected star-sized, impact-produced dust clump in the terrestrial zone \citep{Su2022}. These systems demonstrate that the observed infrared excess can encode not only the amount of small dust, but also the geometry, temperature distribution, and collisional state of impact debris.

Spectroscopy with the Mid-Infrared Instrument (MIRI) on board the James Webb Space Telescope (JWST) now provides a complementary route to studying EDDs by resolving weak solid-state and gas features that were difficult to isolate in Spitzer-era data. Recent JWST observations of the silica-rich giant-impact candidates HD 172555 and HD 23514 revealed gas associated with evaporating close-in material or volatile-bearing debris, while also showing that mid-infrared dust features can remain comparatively stable over decade-long baselines in some systems \citep{Samland2025,Su2025}. These results motivate a cautious interpretation of EDD variability, since changes in flux or color need not imply a wholesale change in mineralogy. Temperature, emitting area, gas production, and dust composition can evolve on different timescales.

The F3V main-sequence star HD 15407A, at a distance of 49.3 pc \citep{Gaia2023}, hosts one of the most prominent known EDDs \citep{Melis2010}. Early mid-infrared observations revealed a warm debris disk with a fractional luminosity of $\sim 0.005$, well above the $\lesssim10^{-4}$ level expected for steady-state warm debris production in standard collisional-cascade models \citep{Wyatt2007,Wyatt2008}. 
Furthermore, far-infrared photometry from Herschel and AKARI indicated a lack of detectable cold dust beyond $\sim 10$ AU, suggesting that the detected circumstellar emission is dominated by material in the inner planetary region, with little evidence for a substantial cold outer disk \citep{Fujiwara2012b}. Spitzer Infrared Spectrograph (IRS) observations in 2008 detected an abundant population of $\sim 1.0\ \mu\text{m}$ silica dust, implying an origin from the thermally processed outer layers of differentiated rocky bodies \citep{Fujiwara2012a}.

Prior studies of HD 15407A detected the broad solid-state features with Spitzer/IRS, but a single Spitzer epoch could not determine whether later spectral evolution would be compositional, thermal, or geometric. HD 15407A therefore provides a well-constrained target for the JWST-era study of decade-scale EDD evolution. We use the 2023 JWST MIRI/MRS spectra, together with archival Spitzer observations, to establish a 14.95 yr temporal baseline and test whether the spectral evolution requires a new mineral inventory or can be explained mainly by changes in thermal structure and mineral-emitting flux. Section \ref{sec:obs} describes the Spitzer and JWST observations, spectral processing, and modeling framework. Section \ref{sec:results} presents the final no-rim two-isothermal fit and mineralogical results. In Section \ref{sec:discussion}, we discuss the stable mineral inventory, grain-size contribution, flux evolution, and the evolved post-impact debris interpretation. Section \ref{sec:conclusion} summarizes our primary conclusions.

\section{Observations and Data Analysis}\label{sec:obs}
\subsection{Spectral Processing}\label{subsec:spectral_processing}
We used mid-infrared spectra of the HD 15407A debris disk obtained with JWST MIRI/MRS under PID 1206 \citep{Rieke2015,Wells2015,Wright2023}. For the JWST data, we downloaded the uncalibrated MIRI/MRS detector exposures for JWST GTO program 1206, Observation 6, obtained on 2023 September 23, from MAST and processed them with the JWST calibration pipeline version 2.0.1 and CRDS context \texttt{jwst\_1535.pmap}. The observation used four dithers in each of the short, medium, and long MRS grating settings, recorded on the MIRIFUSHORT and MIRIFULONG detectors, giving 24 \texttt{UNCAL} exposures. Following the standard MIRI/MRS point-source workflow, we ran the default \texttt{calwebb\_detector1} (\texttt{Detector1Pipeline}), \texttt{calwebb\_spec2} (\texttt{Spec2Pipeline}), and \texttt{calwebb\_spec3} (\texttt{Spec3Pipeline}) stages. In the Stage 3 run, the pipeline applied outlier detection to the calibrated exposures, built one reconstructed s3d cube for each of the twelve MRS bands, and then ran \texttt{extract\_1d} on these cubes to produce the final observation-level point-source x1d spectra using CRDS reference file \texttt{jwst\_miri\_extract1d\_0004.asdf}. Across the native MIRI/MRS spectrum extracted here, the wavelength-dependent aperture radii are 0.54--2.10 arcsec; band-by-band robust measurements show that these radii are about 1.5--2.2 times the empirical s3d FWHM and that the empirical radial FWHM increases with wavelength. Visual inspection of the reconstructed s3d cube slices shows no obvious nearby contaminating source or structured background component within the aperture. The extraction auto-centers the unresolved source in each cube, applies the residual-fringe and aperture corrections, and retains the absolute pipeline flux calibration.

We stitched the twelve independently extracted x1d spectra, covering MRS channels 1--4 in the short, medium, and long grating bands. The individual band spectra were placed on their native wavelength grids, concatenated without additional scaling, sorted by wavelength, and filtered to remove non-finite points and points with non-positive uncertainties. The resulting native stitched spectrum covers 4.90--28.62~$\mu$m; over 4.90--25.00~$\mu$m, it matches the current MAST Level 3 x1d products with a median flux ratio of 0.9995 and a relative RMS difference of 0.5\%. To maintain consistency with other archive-based analyses, we adopted the current MAST Level 3 x1d products for the subsequent analysis.

To assess temporal variations, we incorporated the historical Spitzer/IRS spectrum of HD 15407A, observed on 2008 October 9 with AOR 26122496 \citep{Houck2004}. We used the merged low-resolution point-source spectrum available as \dataset[Spitzer IRS Enhanced Products]{\doi{10.26131/IRSA399}}. These spectra are produced from final SSC pipeline v18.18 \texttt{bksub.tbl} SL and LL spectra extracted from nod-differenced background-subtracted basic calibrated data with an aperture that expands linearly with wavelength, so their calibrated fluxes are point-source products. For HD 15407A, the product merges eight background-subtracted spectra and reports extraction-profile FWHM values of 2.42, 2.65, 5.96, and 8.15 arcsec for SL2, SL1, LL2, and LL1, respectively. Other available Spitzer products differ modestly from the adopted Spitzer spectrum: the CASSISjuice products \citep{Lebouteiller2011} have median 6--25~$\mu$m flux offsets of +3.2\% for the PSF-weighted low-resolution optimal extraction and +0.1\%/-0.3\% for tapered-column extractions calibrated for partially extended and point-like sources, respectively.

\begin{figure*}[t!]
\centering
\includegraphics[width=0.90\textwidth,height=0.62\textheight,keepaspectratio]{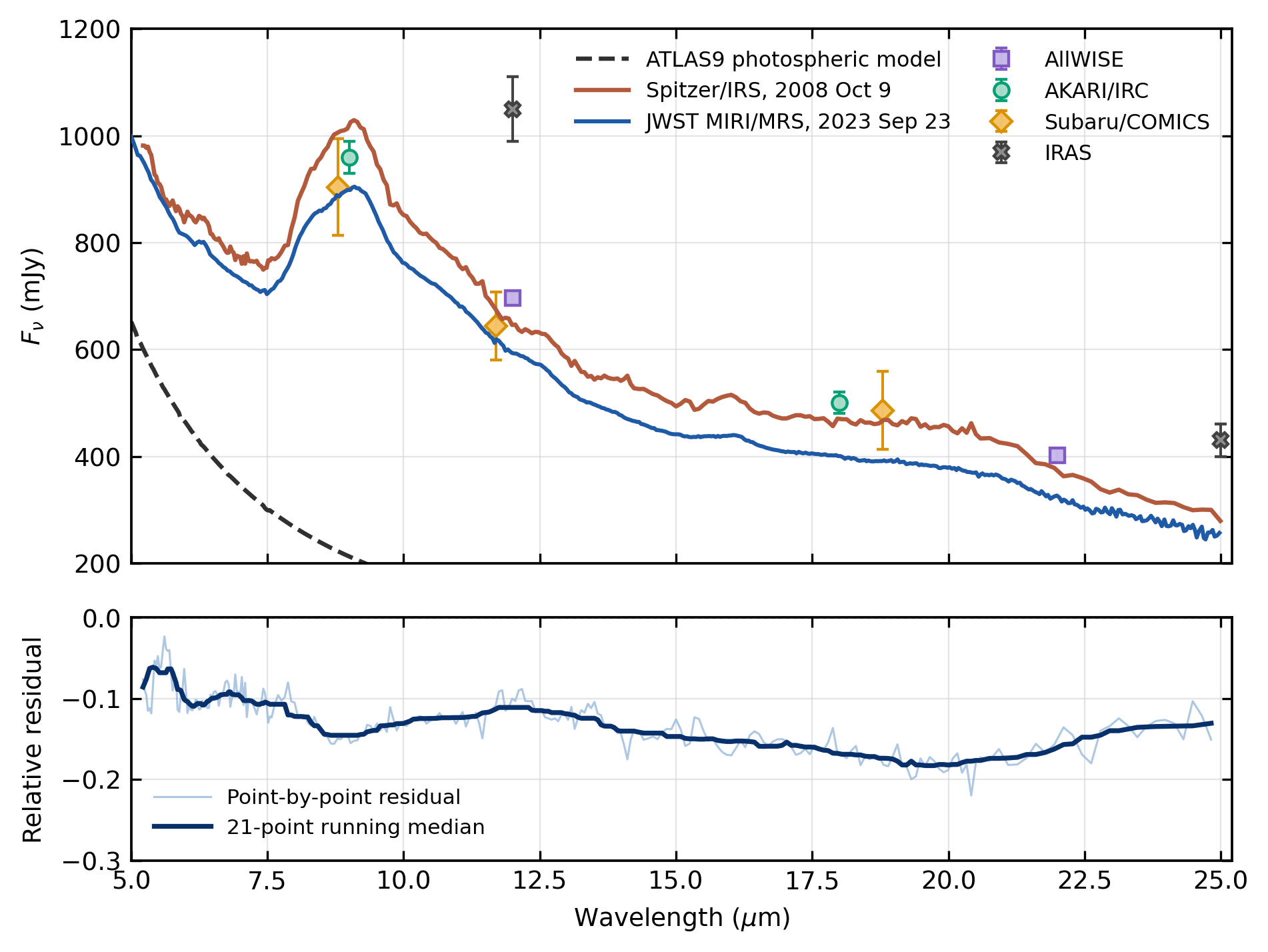}
\caption{Direct comparison of the Spitzer/IRS and JWST MIRI/MRS spectra of HD 15407A. The upper panel is plotted on a linear mJy scale over 5--25~$\mu$m to emphasize the spectral morphology; the black dashed curve shows the adopted ATLAS9 photospheric model from the Castelli--Kurucz grid. The photometry includes the literature AKARI/IRC, Subaru/COMICS, and IRAS mid-infrared measurements compiled by \citet{Fujiwara2012b}, and AllWISE W3/W4 points. The lower panel shows the relative residual, $(F_{\nu,\rm JWST}-F_{\nu,\star})/(F_{\nu,\rm Spitzer}-F_{\nu,\star})-1$, after interpolating the JWST spectrum onto the Spitzer wavelength grid from 5.22 to 25~$\mu$m; the light blue curve is the point-by-point residual and the dark blue curve is a 21-point running median, corresponding to a 0.60--3.39~$\mu$m window over 6--25~$\mu$m.}
\label{fig:epoch_comparison}
\end{figure*}

The Spitzer spectrum is extracted with a slit aperture that expands with wavelength, while the JWST MIRI/MRS spectrum is extracted from IFU cubes with a wavelength-dependent aperture and point-source aperture correction. Both spectra are point-source-calibrated total-flux products for an unresolved target. We treat the absolute cross-instrument comparison by propagating the formal spectral errors together with a calibration-limited term based on the MIRI/MRS point-source flux calibration of \citet{2025AJ....169...67L}. The direct MIRI/MRS--Spitzer/IRS comparison in that work motivates a wavelength-dependent relative calibration floor of 1\% over 6--18~$\mu$m and 3\% over 18--25~$\mu$m. Figure~\ref{fig:epoch_comparison} directly compares the JWST and Spitzer spectra and shows that the merged Spitzer spectrum begins at 5.22~$\mu$m. In the 5.22--6.0~$\mu$m overlap, the Spitzer blue-edge points deviate mildly from the smooth continuum-like slope traced by JWST, while neither spectrum shows a broad astrophysical dust feature in this interval. We attribute this behavior primarily to blue-edge and stitching systematics and adopt a conservative common 6.0--25.0~$\mu$m interval for all dust-model fitting.

\begin{figure*}[t!]
\centering
\includegraphics[width=0.90\textwidth,height=0.62\textheight,keepaspectratio]{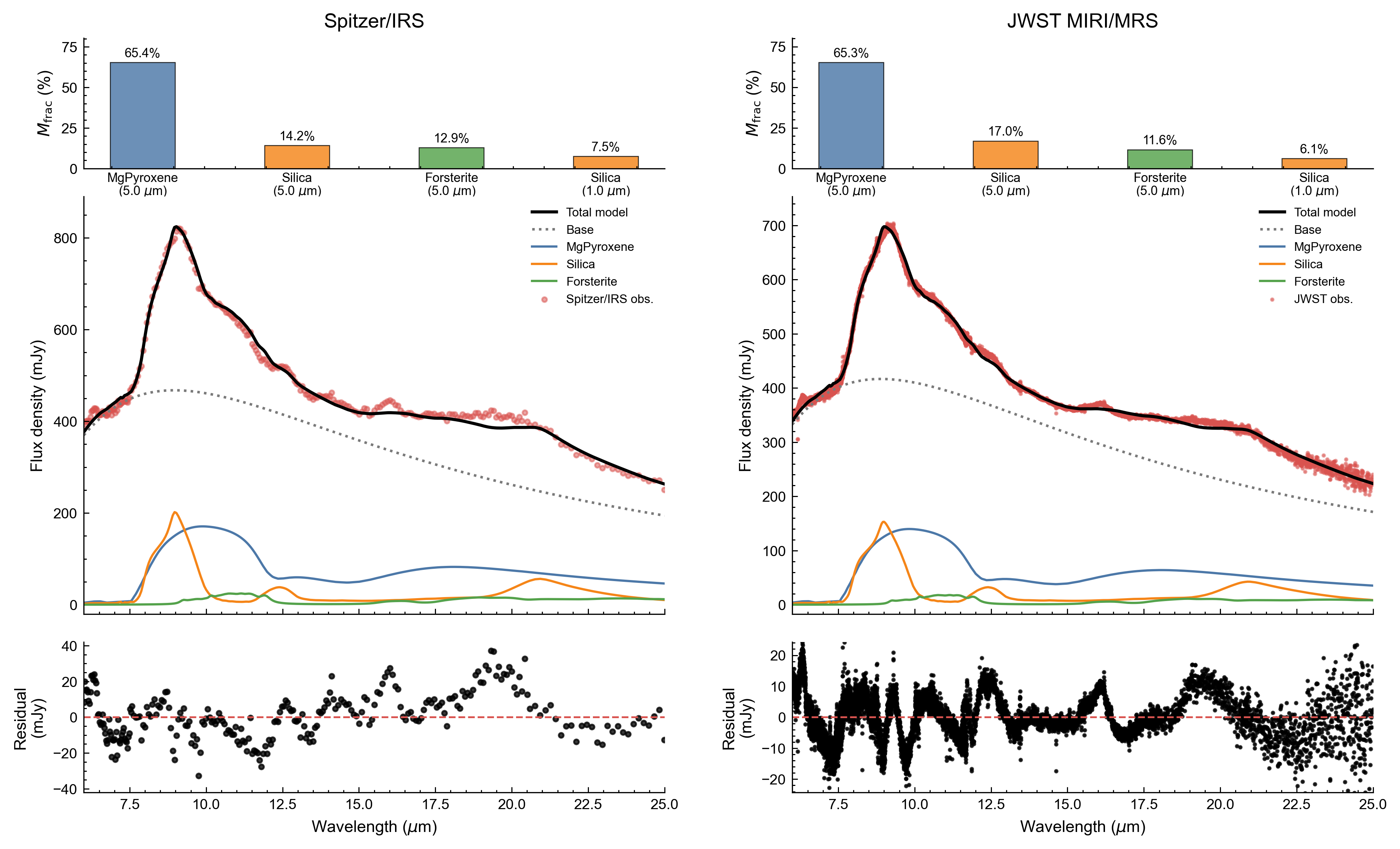}
\caption{No-rim DuCK fits to the Spitzer spectrum and the JWST MIRI/MRS spectrum. The fits use the same fixed ATLAS9 photospheric model from the Castelli--Kurucz grid and are shown after photosphere subtraction. Both epochs are fitted with the same reduced mineral basis: Mg-rich pyroxene, silica, and forsterite. The upper panels show the normalized fitted mineral weights, the middle panels show the model decomposition, and the lower panels show residuals.}
\label{fig:combined_fits}
\end{figure*}

\subsection{Model Selection and Mineralogical Modeling}\label{subsec:modeling}
We analyzed the spectra with a modified version of the DuCK/DuCKLinG dust-emission framework \citep{Kaeufer2024}. DuCK separates the observed spectrum into a fixed stellar photosphere, optically thin mineral emission, and smooth continuum components that describe the underlying thermal structure. Because HD 15407A is a gas-poor debris disk rather than an optically thick protoplanetary disk, we adopt a no-rim two-isothermal model as the baseline. Here ``two-isothermal'' means that the fitted dust emission has two single-temperature Planck components, $T_{\rm surf}$ and $T_{\rm base}$; because of degeneracies among temperature, emitting area, grain opacity, and continuum decomposition, these components need not correspond to two spatially resolved rings and may instead represent different grain populations or effective thermal layers within the same unresolved debris reservoir. DuCK can also evaluate power-law temperature structures for radially distributed dust, but the final HD 15407A fits support single-temperature components for both the base continuum and the mineral-emitting surface.

For the fixed stellar component, we adopted an ATLAS9 stellar atmosphere model from the Castelli--Kurucz grid \citep{CastelliKurucz2003}, with the literature parameters $T_{\rm eff}=6500$ K, $\log g=4.0$, and $[\mathrm{M}/\mathrm{H}]=0$ \citep{Melis2010,Fujiwara2012a}, together with $A_V=0$ and the Gaia DR3 distance $d=49.339$ pc. We fitted only the scale $(R_\star/d)^2$ to Gaia DR3 $G_{\rm BP}$, $G$, and $G_{\rm RP}$ and 2MASS $J$, $H$, and $K_s$ photometry \citep{Gaia2023,Skrutskie2006}, excluding longer-wavelength bands affected by the warm-dust excess. The fitted scale gives $R_\star=1.40\pm0.01~R_\odot$ and $L_\star=3.15\pm0.03~L_\odot$. An independent multiband \href{https://github.com/pjchen3/sedforge}{SEDforge} fit, with $T_{\rm eff}$ as the only free atmospheric parameter, gives $6516\pm38$ K, consistent with the adopted value. We used the same photospheric model for both epochs and linearly interpolated its emergent flux density onto each observed wavelength grid. Before this interpolation, sparsely sampled intervals in the native model wavelength grid longward of 10~$\mu$m were densified in $\log\lambda$--$\log F_\nu$ space.
For the dust opacities, we first used eight standard mineral species: Mg-rich olivine, Mg-rich pyroxene, olivine, pyroxene, silica, enstatite, forsterite, and fayalite. The Mg-rich labels refer to the Fe-poor magnesium end-member amorphous templates in the DuCK library, i.e., Mg/(Mg+Fe)=1 in the template naming, whereas the generic olivine and pyroxene templates represent Mg--Fe silicate glasses. Among these, enstatite, forsterite, and fayalite are crystalline minerals. Each species was evaluated at grain radii of 0.1, 1.0, and 5.0~$\mu$m, representing small, intermediate, and relatively large grains within the mid-infrared emitting population. We used the pre-computed absorption-efficiency curves, $Q_{\rm abs}(\nu)$, for Gaussian random field (GRF) particle shapes available in the DuCK opacity library to approximate compact randomly irregular mineral grains \citep{Grynko2013,Grynko2014}. As a controlled opacity-shape check, otherwise identical fits using the distribution of hollow spheres (DHS) prescription showed that the joint Spitzer and JWST spectra, driven by the JWST likelihood, favor the adopted GRF opacity branch rather than DHS. For these $Q_{\rm abs}$ templates, the mass absorption coefficient is expressed as $\kappa_\nu=3Q_{\rm abs}(\nu)/(4\rho a)$, where $a$ is the grain radius in cm and $\rho$ is the material density supplied with each template; the opacity curves themselves are not renormalized in the fit. The optical constants in the adopted DuCK opacity library draw mainly on \citet{Dorschner1995} for amorphous Mg--Fe olivine and pyroxene glasses, \citet{Henning1997} for silica and related dust analogues, and \citet{Fabian2001} for crystalline olivines including forsterite and fayalite, with enstatite taken from the crystalline-silicate part of the same library.

The photosphere-subtracted residual spectrum in Figure~\ref{fig:epoch_comparison} is dominated by a broad flux-level offset and does not show a new feature-specific residual that would require an epoch-dependent mineral basis. After the initial eight-mineral screen, we retained the union of species whose normalized fitted mineral weights exceeded 5\% in at least one epoch so that both epochs could be refit and compared with the same dominant mineral carriers. This 5\% cutoff is a pragmatic screening threshold rather than a formal detection limit; within the fixed model the formal posterior intervals are small, but sub-5\% components are more vulnerable to degeneracies among weak opacity features, continuum placement, and grain-size weighting. For HD 15407A this selection retained Mg-rich pyroxene, silica, and forsterite. We then refit the Spitzer and JWST spectra with this common three-mineral basis and the same mineral and grain-size templates in both epochs, providing a uniform comparison of thermal structure, emitting area, and relative mineral weights between epochs. The reported mineral weights are normalized fitted emitting-mass weights, $f_{j,a}=M_{j,a}/\sum_{j,a}M_{j,a}$, within the adopted opacity library and grain-size grid.

The adopted no-rim two-isothermal model is
\begin{equation}
F_\nu^{\rm mod} =
F_{\nu,\star}
+ C_{\rm base}\,B_\nu(T_{\rm base})
+ \frac{B_\nu(T_{\rm surf})}{d^2}
   \sum_{j,a} M_{j,a}\,\kappa_{j,a}(\nu),
\label{eq:hd15407a_baseline_model}
\end{equation}
where $F_{\nu,\star}$ is the fixed stellar photosphere, $C_{\rm base}$ is the apparent solid-angle normalization of the smooth base continuum, $M_{j,a}$ is the optically thin emitting mass for mineral species $j$ and grain radius $a$, and $\kappa_{j,a}$ is the corresponding mass absorption coefficient. The nonlinear parameters are $T_{\rm surf}$ and $T_{\rm base}$, for which we adopted broad uniform priors of 100--1500 K; for each sampled temperature pair, the non-negative linear amplitudes $C_{\rm base}$ and $M_{j,a}$ were solved by non-negative least squares (NNLS). The likelihood assumes independent Gaussian errors, using the point-by-point $1\sigma$ uncertainties provided with the Spitzer spectrum and the pipeline-provided uncertainties of the JWST x1d spectra. When reporting absolute masses, we convert the fitted NNLS template amplitudes into physical emitting masses using the adopted GRF opacity templates, grain-size grid, and distance. These masses depend on the absolute $\kappa_\nu$ scale of the opacity templates, grain-size grid, surface temperature, and continuum decomposition, and are used only for the fitted optically thin surface component. The reported fitting uncertainties are formal statistical uncertainties for the adopted fixed model and do not include opacity, grain-size, continuum-decomposition, or calibration systematics.

\section{Results}\label{sec:results}

After subtracting the fixed stellar photosphere, the integrated 6--25~$\mu$m dust-excess fluxes on the Figure~\ref{fig:epoch_comparison} grid are $8.678\pm0.006$ Jy~$\mu$m for Spitzer/IRS and $7.466\pm0.002$ Jy~$\mu$m for JWST MIRI/MRS. The corresponding formal JWST/Spitzer ratio is $0.860\pm0.001$. Accounting for the relative calibration uncertainty (1\% over 6--18~$\mu$m and 3\% over 18--25~$\mu$m) gives an adopted decrease of $14.0\pm1.2\%$, significant at $11.9\sigma$. The residual panel in Figure~\ref{fig:epoch_comparison} gives a consistent result: the 6--25~$\mu$m median residual is $-13.1\%$, and the 21-point running median, corresponding to a 0.60--3.39~$\mu$m window on the Spitzer grid over this interval, spans $-18.3\%$ to $-9.2\%$. The central wavelength and morphology of the main solid-state feature remain broadly similar.

Figure~\ref{fig:epoch_comparison} shows the mid-infrared photometry, including literature AKARI/IRC, Subaru/COMICS, and IRAS points compiled by \citet{Fujiwara2012b} and the AllWISE W3 and W4 photometry \citep{Cutri2013}. The IRAS PSC photometry \citep{IRASExplanatory1988} comes from 1983 and has three hours-confirmed coverages; the AKARI/IRC S9W and L18W measurements \citep{Ishihara2010} were obtained in two survey visits on 2006 August 17--18 and 2007 February 15--16; the Subaru/COMICS N8.8, N11.7, and Q18.8 archive frames span 2007 July 5--7 and 2008 July 17--18; and the AllWISE W3 and W4 points come from 15 exposures on 2010 February 8--9. Combined with the Spitzer and JWST spectra, these archival points trace a long-term decline in the mid-infrared excess from the IRAS era through Spitzer to JWST; we discuss this heterogeneous time-domain context in Section~\ref{sec:discussion}.

Figure~\ref{fig:combined_fits} shows the stellar-subtracted model decompositions and residuals for the final no-rim fits. Table~\ref{tab:free_results} lists the final no-rim two-isothermal fit, including the fitted continuum parameters, model-dependent optically thin emitting mass, relative residual RMS, and normalized fitted mineral weights. The relative residual RMS is computed from $(F_{\rm obs}-F_{\rm mod})/F_{\rm obs}$ over the fitted wavelength range. The Spitzer epoch is fitted with $T_{\rm surf}=695.8^{+7.3}_{-7.6}$ K and $T_{\rm base}=569.6^{+1.0}_{-1.2}$ K. The JWST epoch is fitted with $T_{\rm surf}=754.8^{+0.9}_{-1.0}$ K and $T_{\rm base}=574.1^{+0.1}_{-0.1}$ K. These intervals are formal 16th--84th percentile posterior intervals for the adopted empirical model, with the corresponding fixed-model posterior corner plots shown in Appendix Figures~\ref{fig:table1_posterior_spitzer} and \ref{fig:table1_posterior_jwst}. Under the fixed Mg-rich pyroxene, silica, and forsterite model, the normalized component weights shift by $-1.4$ to $+2.8$ percentage points. Both epochs retain the same mineral basis, with no need for a new compositionally distinct carrier. Because the normalized weights, $C_{\rm base}$, and optically thin emitting mass are correlated and model-dependent, we interpret the percent-level weight shifts as fixed-model redistribution among the same carriers rather than as a robust compositional transformation. The mid-infrared flux level changes modestly, while the dominant solid-state carriers remain the same.

\begin{deluxetable*}{lccc}
\tablewidth{0pt}
\tablecaption{Disk and Mineralogical Properties of HD 15407A \label{tab:free_results}}
\tablehead{
\colhead{Quantity} & \colhead{Spitzer} & \colhead{JWST MIRI/MRS} & \colhead{Descriptive change}
}
\startdata
Observation date & 2008 Oct 9 & 2023 Sep 23 & 14.95 yr \\
$T_{\rm surf}$ (K) & $695.8^{+7.3}_{-7.6}$ & $754.8^{+0.9}_{-1.0}$ & $+58.9$ K \\
$T_{\rm base}$ (K) & $569.6^{+1.0}_{-1.2}$ & $574.1^{+0.1}_{-0.1}$ & $+4.5$ K \\
Base-continuum coefficient $C_{\rm base}$ (sr) & $(1.34^{+0.01}_{-0.01})\times10^{-16}$ & $(1.16^{+0.01}_{-0.01})\times10^{-16}$ & $-13.0\%$ \\
Fitted optically thin surface-component mass ($M_\oplus$) & $(1.29^{+0.03}_{-0.03})\times10^{-7}$ & $(8.79^{+0.03}_{-0.02})\times10^{-8}$ & $-32.1\%$ \\
Relative residual RMS & 2.30\% & 1.34\% & \nodata \\
Mg-rich pyroxene weight, 5.0~$\mu$m & $65.4^{+0.7}_{-0.6}\%$ & $65.3^{+0.1}_{-0.1}\%$ & $-0.1\%$ \\
Silica weight, 5.0~$\mu$m & $14.2^{+0.1}_{-0.1}\%$ & $17.0^{+0.1}_{-0.1}\%$ & $+2.8\%$ \\
Forsterite weight, 5.0~$\mu$m & $12.9^{+0.6}_{-0.6}\%$ & $11.6^{+0.1}_{-0.1}\%$ & $-1.3\%$ \\
Silica weight, 1.0~$\mu$m & $7.5^{+0.1}_{-0.1}\%$ & $6.1^{+0.1}_{-0.1}\%$ & $-1.4\%$ \\
Total 5.0~$\mu$m weights & $92.5^{+0.1}_{-0.1}\%$ & $93.9^{+0.1}_{-0.1}\%$ & $+1.4\%$ \\
\enddata
\tablecomments{The listed masses are model-dependent estimates for the fitted optically thin mineral-emitting surface component, computed with the adopted GRF opacity templates, grain-size grid, and distance. They should not be interpreted as total disk masses. Percentages are normalized fitted mineral weights. The fitted optically thin surface-component mass is the sum over the retained mineral species and grain-size terms in this fixed basis; species- or size-specific fitted masses can be obtained by multiplying this mass by the corresponding normalized fitted weight. Uncertainties are rounded formal 16th--84th percentile posterior intervals for the adopted fixed model. The listed changes for $C_{\rm base}$ and emitting mass are fractional changes relative to the Spitzer values, while mineral-weight changes are percentage-point differences.}
\end{deluxetable*}

\section{Discussion}\label{sec:discussion}
\subsection{Stable Mineralogy and Grain-size Contribution}

A main result of this work is the stability of the solid-state spectral carriers. Both the Spitzer and JWST spectra are reproduced with the same reduced opacity basis, consisting of Mg-rich pyroxene, silica, and forsterite. The normalized fitted mineral weights change only at the level of a few percentage points: Mg-rich pyroxene remains the dominant component, silica provides a persistent secondary contribution, and forsterite is retained as a smaller crystalline component. The spectral evolution between 2008 and 2023 does not require the injection of a compositionally distinct dust population.

\begin{figure*}[t!]
\centering
\includegraphics[width=0.82\textwidth,height=0.42\textheight,keepaspectratio]{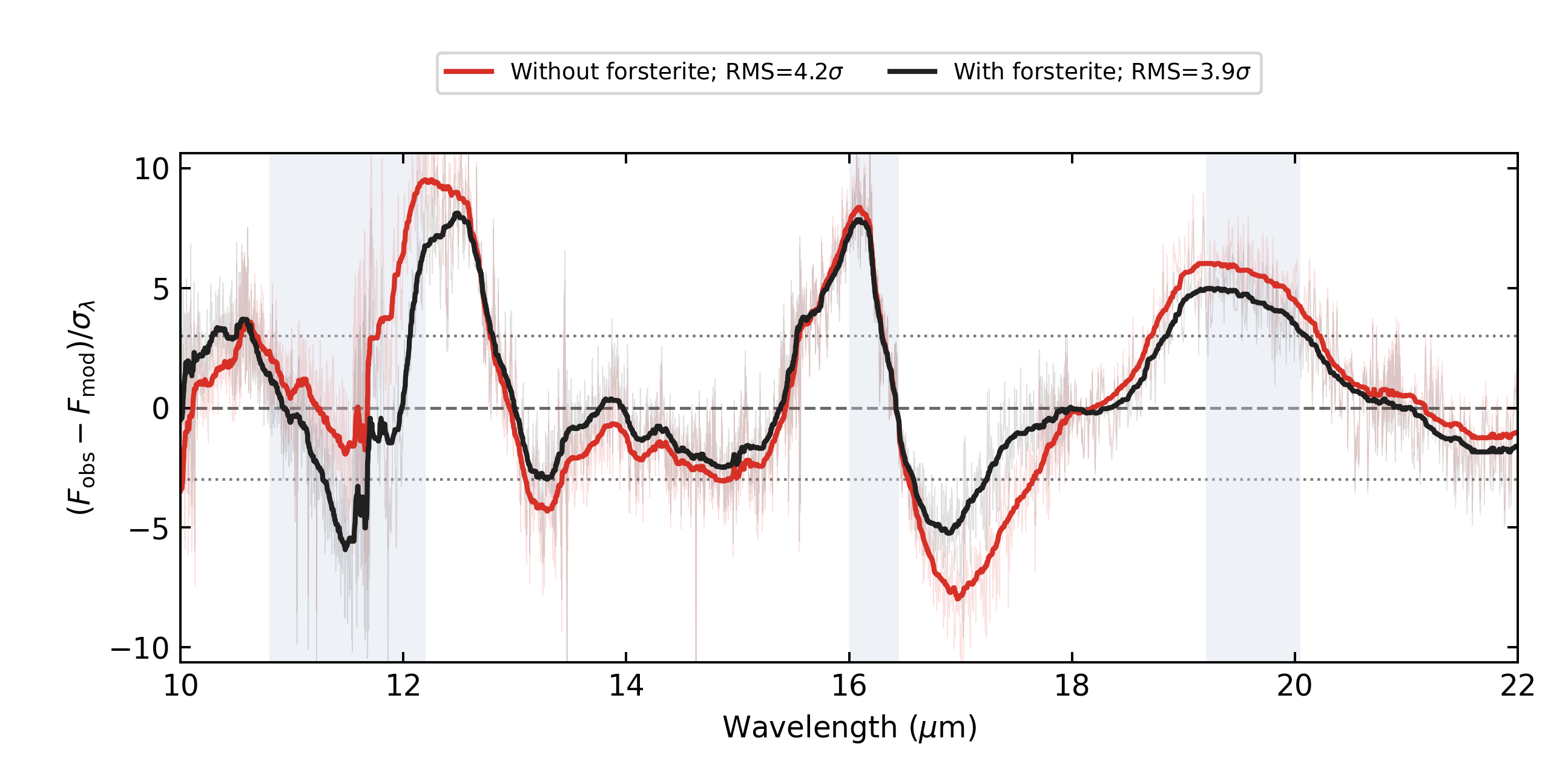}
\caption{JWST-only 10--22~$\mu$m standardized-residual comparison for a controlled no-forsterite ablation test. The red curve shows the residual from an otherwise identical two-mineral fit using only Mg-rich pyroxene and silica, while the black curve shows the adopted three-mineral baseline that includes forsterite. The ordinate is $(F_{\rm obs}-F_{\rm mod})/\sigma_\lambda$, where $\sigma_\lambda$ is the formal flux-density uncertainty at each wavelength. The dashed horizontal line marks zero residual, and the dotted lines mark $\pm3\sigma$. Thin curves show native-grid standardized residuals and thick curves show 101-point running medians for display; the RMS values in the legend are computed over the native 6--25~$\mu$m grid. Shaded bands mark representative forsterite-sensitive wavelength regions.}
\label{fig:forsterite_ablation}
\end{figure*}

The no-forsterite ablation tests clarify the role of this smaller crystalline component. Because \citet{Fujiwara2012a} fitted the Spitzer spectrum without forsterite, we refitted both spectra with only Mg-rich pyroxene and silica under the same framework. The Spitzer-only ablation does not favor a required forsterite component, consistent with the earlier forsterite-free Spitzer fit. In contrast, the high-S/N, high-resolution JWST MIRI/MRS spectrum supports retaining forsterite within the adopted opacity basis (Figure~\ref{fig:forsterite_ablation}). Across the 11--20~$\mu$m forsterite-sensitive interval, MIRI/MRS has $R\simeq1500$--3300, compared with $R\simeq60$--130 for low-resolution Spitzer/IRS, corresponding to a typical resolving-power advantage of about 20--30 \citep{Houck2004,Wells2015,Wright2023}. Removing forsterite increases the flux-density residual RMS from 7.1 to 7.5~mJy over 6--25~$\mu$m (standardized-residual RMS: $3.9\sigma$ to $4.2\sigma$) and from 10.4 to 12.9~mJy over 19.2--20.05~$\mu$m (standardized-residual RMS: $4.5\sigma$ to $5.6\sigma$), while increasing the BIC by $2.25\times10^4$. In the two-mineral fits for both epochs, the removed forsterite contribution is absorbed mainly by Mg-rich pyroxene, which increases to about 78--79\%, while silica remains near 21--22\%. The Spitzer--JWST mineral mixture remains nearly unchanged even without forsterite, so the conclusion of stable mineralogy does not depend on whether this component is included. We retain forsterite in the common three-mineral basis because it is favored by JWST and provides a uniform two-epoch comparison, but we do not claim a unique forsterite identification because part of the improvement could reflect limitations of the adopted opacity library, continuum decomposition, or reduced mineral basis.

The fitted mineral weights are also stable across the adopted grain-size grid. In both epochs, the normalized mineral mixture is dominated by the 5.0~$\mu$m components, which account for 92.5--93.9\% of the fitted mineral weight. Sub-5~$\mu$m contributions are confined to silica and remain minor, primarily through the 1.0~$\mu$m term. This 5.0~$\mu$m preference differs from the earlier 1.5~$\mu$m mineral interpretation of \citet{Fujiwara2012a}: their silicate search used amorphous olivine/pyroxene sizes only up to 2.0~$\mu$m and selected 1.5~$\mu$m amorphous pyroxene, whereas our weights are defined on a 0.1, 1.0, and 5.0~$\mu$m template grid. The different available grain-size grids are therefore likely a major source of the apparent size difference. For context, the adopted atmosphere normalization gives $L_\star=3.15~L_\odot$. For a spherical grain, the radiation-pressure-to-gravity ratio is 
\begin{equation}
\beta(a)\equiv\frac{F_{\rm rad}}{F_{\rm grav}}
 = \frac{3L_\star\langle Q_{\rm pr}\rangle_\star}
 {16\pi cGM_\star\rho a}.
\label{eq:blowout_beta}
\end{equation}
Here $\langle Q_{\rm pr}\rangle_\star$ is the radiation-pressure efficiency averaged over the stellar spectrum. Setting $\beta=0.5$ defines the blowout radius,
\begin{equation}
a_{\rm blow}
 = \frac{3L_\star\langle Q_{\rm pr}\rangle_\star}
 {8\pi cGM_\star\rho}
 = 1.15~\mu{\rm m}
 \left(\frac{L_\star}{L_\odot}\right)
 \left(\frac{M_\star}{M_\odot}\right)^{-1}
 \left(\frac{\rho}{1~{\rm g~cm^{-3}}}\right)^{-1}
 \langle Q_{\rm pr}\rangle_\star .
\label{eq:blowout_radius}
\end{equation}
Adopting $M_\star=1.4~M_\odot$, $\rho=3.0~{\rm g~cm^{-3}}$, and $\langle Q_{\rm pr}\rangle_\star=1$ gives $a_{\rm blow}=0.86~\mu$m \citep{Burns1979}. Compared with HD 23514, whose mid-infrared spectrum requires abundant 0.1--0.5~$\mu$m silica- and silicate-like grains \citep{Su2025}, the weak sub-5~$\mu$m weights in HD 15407A suggest a limited recent supply of very small grains.

The thermal decomposition also favors evolution in component flux rather than composition. The fitted surface layer becomes hotter, from $695.8^{+7.3}_{-7.6}$ K to $754.8^{+0.9}_{-1.0}$ K, while the base-continuum color temperature remains nearly unchanged at $569.6^{+1.0}_{-1.2}$--$574.1^{+0.1}_{-0.1}$ K. Within this fixed decomposition, the fitted $C_{\rm base}$ coefficient decreases by 13.0\%. The epoch-to-epoch difference is better described as a redistribution of thermal structure and component flux than as a reset of the underlying dust composition. The stable mineral inventory and similar grain-size contribution indicate that HD 15407A was not observed in two chemically distinct dust states; the spectra instead trace the slow evolution of an existing rocky debris reservoir.

\subsection{Mineral-emitting Flux and Effective Decay Timescale}

We compare the two spectral epochs through three nested quantities: the fitted surface-layer mass, the fitted surface-component flux, and the stellar-subtracted total excess. The optically thin emitting masses listed in Table~\ref{tab:free_results} refer to the model-dependent mineral-emitting surface component. The fitted surface dust mass decreases from $1.29\times10^{-7}$ to $8.79\times10^{-8}~M_\oplus$, a ratio of 0.679. This is the most model-dependent quantity because the mass coefficient is inferred from $F_\nu \simeq M_{\rm dust}\kappa_\nu B_\nu(T)/d^2$ after adopting the fitted surface temperature, wavelength-dependent opacity mixture, and base/surface continuum decomposition. This mass change must therefore be interpreted together with the simultaneous 8\% rise in $T_{\rm surf}$.

The fitted 6.0--25.0~$\mu$m flux of the same mineral-emitting surface component gives a less model-dependent radiative measure, with $F_{\rm surf,2023}/F_{\rm surf,2008}=0.789$. For fixed opacity and temperature this would correspond to a comparable decrease in warm mineral-emitting surface area. The stellar-subtracted total 6.0--25.0~$\mu$m excess decreases more modestly, by 14.0\%, or a ratio of 0.860. The sequence of ratios---0.679 for the fitted surface mass, 0.789 for the surface-component flux, and 0.860 for the total excess---shows that the strongest apparent change occurs in the model-dependent mineral-emitting surface layer, while the integrated disk excess evolves more slowly.

\begin{figure*}[t!]
\centering
\includegraphics[width=0.82\textwidth,height=0.45\textheight,keepaspectratio]{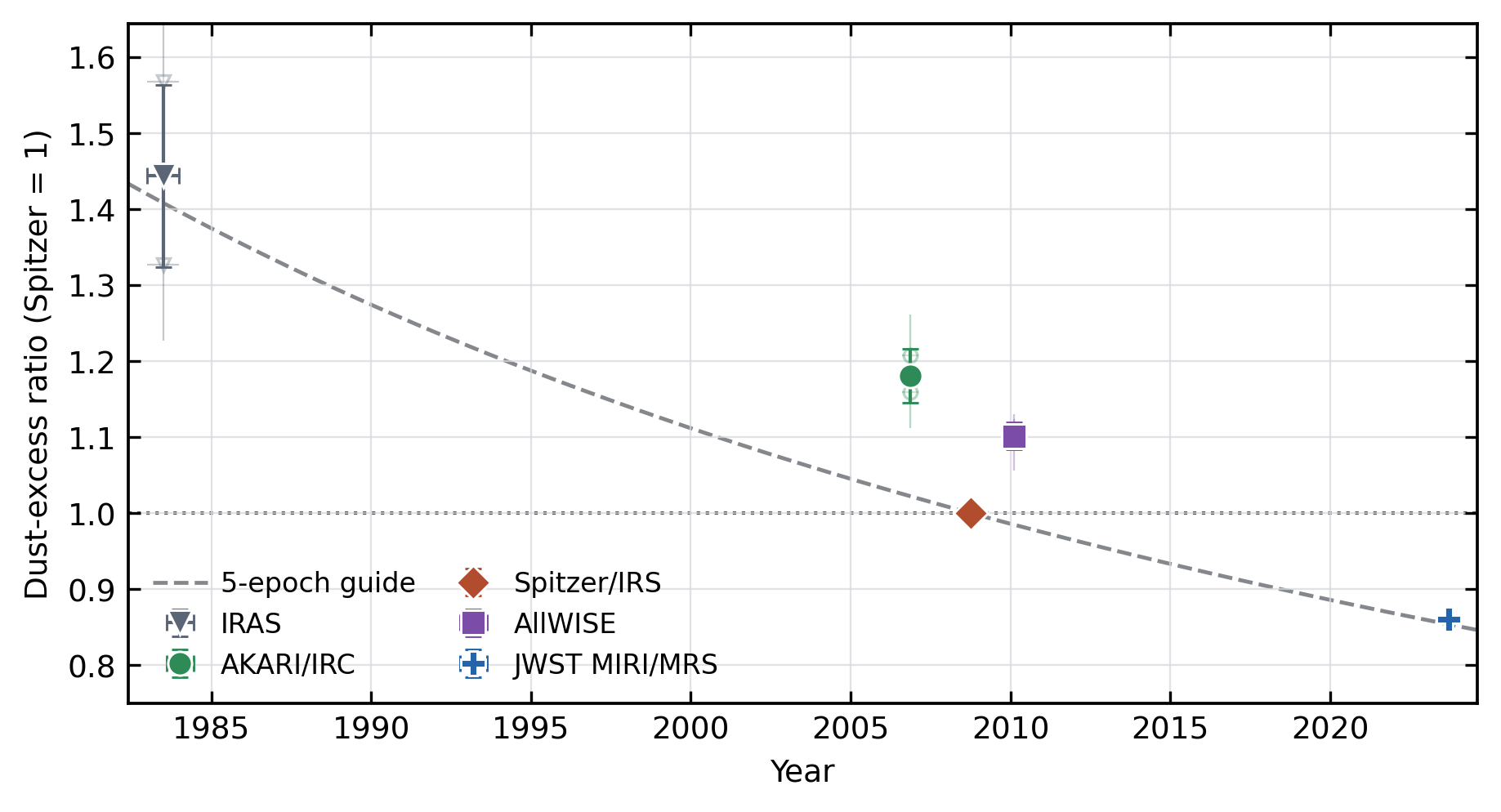}
\caption{Multi-epoch mid-infrared dust-excess ratios for HD 15407A, normalized to the Spitzer/IRS epoch. IRAS, AKARI/IRC, and AllWISE use stellar-subtracted broad-band photometry; the JWST MIRI/MRS point is the direct stellar-subtracted 6.0--25.0~$\mu$m spectral-integral ratio. Small transparent symbols show individual archival broad-band ratios, and filled symbols show combined epoch values. The dashed curve is the collisionally motivated five-epoch guide, $R(t)=[1+(t-t_{\rm Spitzer})/t_c]^{-1}$, with $t_c=87^{+17}_{-12}$ yr.}
\label{fig:epoch_timeseries}
\end{figure*}

To place this two-epoch spectral result in a longer time-domain context, we constructed a multi-epoch mid-infrared dust-excess time series (Figure~\ref{fig:epoch_timeseries}). For each archival broad-band photometric point, we first subtracted the bandpass-integrated stellar photosphere and divided the resulting dust-excess flux by the Spitzer dust-excess flux in the same band. The Spitzer reference was computed by convolving the Spitzer spectrum with the corresponding IRAS 12 and 25~$\mu$m, AKARI/IRC S9W and L18W, or WISE W3 and W4 response curve and subtracting the same stellar photosphere, so the Spitzer ratio is unity by construction. For JWST, we used the direct stellar-subtracted 6.0--25.0~$\mu$m spectral-integral ratio relative to Spitzer, matching the two-epoch spectral comparison above. The resulting representative ratios are $1.44\pm0.12$ for IRAS, $1.18\pm0.04$ for AKARI/IRC, $1.101\pm0.018$ for AllWISE, 1.000 for Spitzer/IRS, and $0.860\pm0.001$ for JWST MIRI/MRS. The Subaru/COMICS N8.8, N11.7, and Q18.8 points, analyzed in the same way, give a combined dust-excess ratio of $0.93\pm0.07$; because this uncertainty is comparatively large, they provide only a weak constraint on the dust-excess time series and are not included in the fit.

Although the archival photometric constraints are heterogeneous, their long temporal baseline still constrains the secular mid-infrared trend. Motivated by the 5.0~$\mu$m-dominated mineral weights, which are consistent with a limited recent supply of very small grains, we fit the five long-baseline epoch values with a simple collisional-cascade guide. The guide, $R(t)=[1+(t-t_{\rm Spitzer})/t_c]^{-1}$ with $R=1$ at the Spitzer epoch \citep{Wyatt2007,Wyatt2008}, gives $t_c=87^{+17}_{-12}$ yr after adding 5\% systematic uncertainties to the IRAS, AKARI/IRC, and AllWISE photometric epochs and 3\% to the Spitzer/IRS and JWST MIRI/MRS spectral epochs. The parameter $t_c$ is the characteristic fading time of this guide: after one $t_c$, the model dust-excess ratio falls from the Spitzer-normalized value of 1 to 0.5. Under this interpretation, an HD 15407A-like debris reservoir with limited recent supply of very small grains would halve on the fitted $t_c$ timescale and then continue fading over several hundred years, dropping substantially below its current extreme mid-infrared excess level. This fitted timescale is an effective decay timescale for the observed mineral-emitting flux, not the physical lifetime of individual grains; individual micron-sized grains may be removed faster, while ongoing collisional production from larger fragments can maintain the slowly fading observable reservoir.

\subsection{Implications for an Impact-generated Debris Reservoir}

HD 15407A has long been interpreted as a prominent warm EDD. Its unusually large fractional infrared luminosity, prominent silica-rich mid-infrared spectrum, and lack of a substantial cold outer-disk component have motivated scenarios in which the observed dust is linked to a recent giant impact during rocky planet formation \citep{Melis2010,Fujiwara2012b,Fujiwara2012a}. In such a scenario, a giant impact can initially generate vapor, melt droplets, and abundant fine silicate dust. The smallest grains are then rapidly removed by radiation pressure and destructive mutual collisions, while larger fragments continue to collisionally grind down and resupply the observable dust. This production may reflect the decay of a single major event, or repeated collisions within the same dynamically excited rocky reservoir.

Recent JWST--Spitzer comparisons show that EDD spectral evolution is diverse. HD 23514 preserves a prominent silica feature over a 15 yr Spitzer--JWST baseline and shows no large-scale ($>30\%$) change over the longer IRAS-to-JWST interval, even though its submicron silica and volatile gas require special conditions to avoid rapid removal or photodissociation \citep{Su2025}. In $\beta$~Pic, the 5--15~$\mu$m continuum excess decreased between Spitzer and JWST, the 10~$\mu$m silicate profile changed, and the 18 and 23~$\mu$m crystalline-forsterite features nearly disappeared, plausibly because radiation pressure removed small Rayleigh-limit crystalline grains over the intervening $\sim20$ yr \citep{Chen2024}. HD 15407A differs from both systems: it preserves the main mineral fingerprint, lacks evidence for the abundant 0.1--0.5~$\mu$m grains inferred in HD 23514, and does not show the near-disappearance of Rayleigh-limit crystalline-grain features inferred for $\beta$~Pic. These comparisons identify HD 15407A as a candidate warm debris disk with limited recent supply of very small grains and a fading mid-infrared excess. Here, this means that the recent input of very small grains from a new major impact, external delivery of dust-producing material, or efficient collisional grinding of larger fragments is very limited.

The Spitzer--JWST comparison adds two constraints on this interpretation. The spectra do not show evidence for a major mineralogical transition between 2008 and 2023. They preserve the same Mg-rich pyroxene--silica--forsterite mineral basis and a similar 5.0~$\mu$m-dominated grain-size contribution. These properties suggest that by the time of the 2008 Spitzer observation, HD 15407A had likely already evolved beyond the most rapidly changing post-impact phase dominated by freshly produced vapor and the shortest-lived small grains. The system is consistent with an evolved phase of less efficient collisional replenishment, in which broadly similar micron-sized dust is still produced while the observable mineral-emitting flux declines. The main temporal evolution is the gradual decline and redistribution of warm mineral emission, rather than the appearance of new minerals. The IRAS-to-JWST multi-epoch fit places this observable fading on a decades-to-century scale, with the strongest change associated with the mineral-emitting surface component. If the current trend continues and no substantial new material is supplied, the observable warm emission would be expected to fade significantly over several hundred years.

HD 15407A provides a time-domain example of post-impact debris evolution in which the mineralogical memory of rocky parent material is preserved while the warm mineral-emitting flux decreases. Continued mid-infrared monitoring over longer baselines will be essential for distinguishing among a freely decaying post-impact reservoir, episodic replenishment by repeated collisions, and a longer-lived collisionally active rocky debris belt, thereby constraining the evolution of EDDs and the lifetime of observable giant-impact signatures during terrestrial planet assembly.

\section{Conclusion}\label{sec:conclusion}
We have compared the Spitzer/IRS and JWST MIRI/MRS spectra of the warm EDD around HD 15407A over a 14.95 yr baseline. The stellar-subtracted 6--25~$\mu$m excess decreased by $14.0\pm1.2\%$, while the solid-state spectral morphology remains nearly unchanged. After subtraction of the stellar photosphere, both epochs are described by the same no-rim two-isothermal empirical decomposition, consisting of a smooth base continuum and an optically thin mineral-emitting surface layer.

The main result of this work is the stability of the rocky mineral fingerprint. Both epochs are fitted with the same Mg-rich pyroxene--silica--forsterite basis. Mg-rich pyroxene remains dominant at about 65\%, silica contributes about 22--23\%, and forsterite contributes about 12--13\% of the normalized fitted mineral weight. The no-forsterite ablation tests show that this stable-mineralogy conclusion does not depend on whether forsterite is included in the fit. In both epochs, 93--94\% of the fitted mineral weight is carried by 5.0~$\mu$m grains, with only weak contributions from smaller grains. The 15 yr evolution shows no evidence for the injection of a compositionally distinct dust population or a chemical reset of the dust system.

The main temporal evolution is thermal and geometric. The mineral-emitting surface layer becomes hotter, from $695.8^{+7.3}_{-7.6}$ to $754.8^{+0.9}_{-1.0}$ K, while the smooth base component remains nearly fixed at $\simeq572$ K. The IRAS-to-JWST multi-epoch ratios are described by a simple collisional-cascade guide with $t_c=87^{+17}_{-12}$ yr, corresponding to a decline of the Spitzer-normalized dust-excess ratio to 0.5 after one $t_c$. Because the fitted dust mass is coupled to temperature, opacity, and continuum decomposition, this fading is best interpreted primarily as evolution of the warm mineral-emitting surface component rather than as the lifetime of individual grains.

HD 15407A likely entered an evolved post-impact phase before the 2008 Spitzer observation. The system preserves the mineralogical memory of its rocky parent material while its warm mineral-emitting flux is gradually fading over several hundred years, consistent with limited recent supply of very small grains. Continued mid-infrared monitoring will test whether this fading follows a single collisional-cascade decay or whether future observations capture intermittent external replenishment.

\appendix

\section{Formal Posterior Distributions}
\restartappendixnumbering

\begin{figure*}[t!]
\centering
\includegraphics[width=0.95\textwidth,height=0.84\textheight,keepaspectratio]{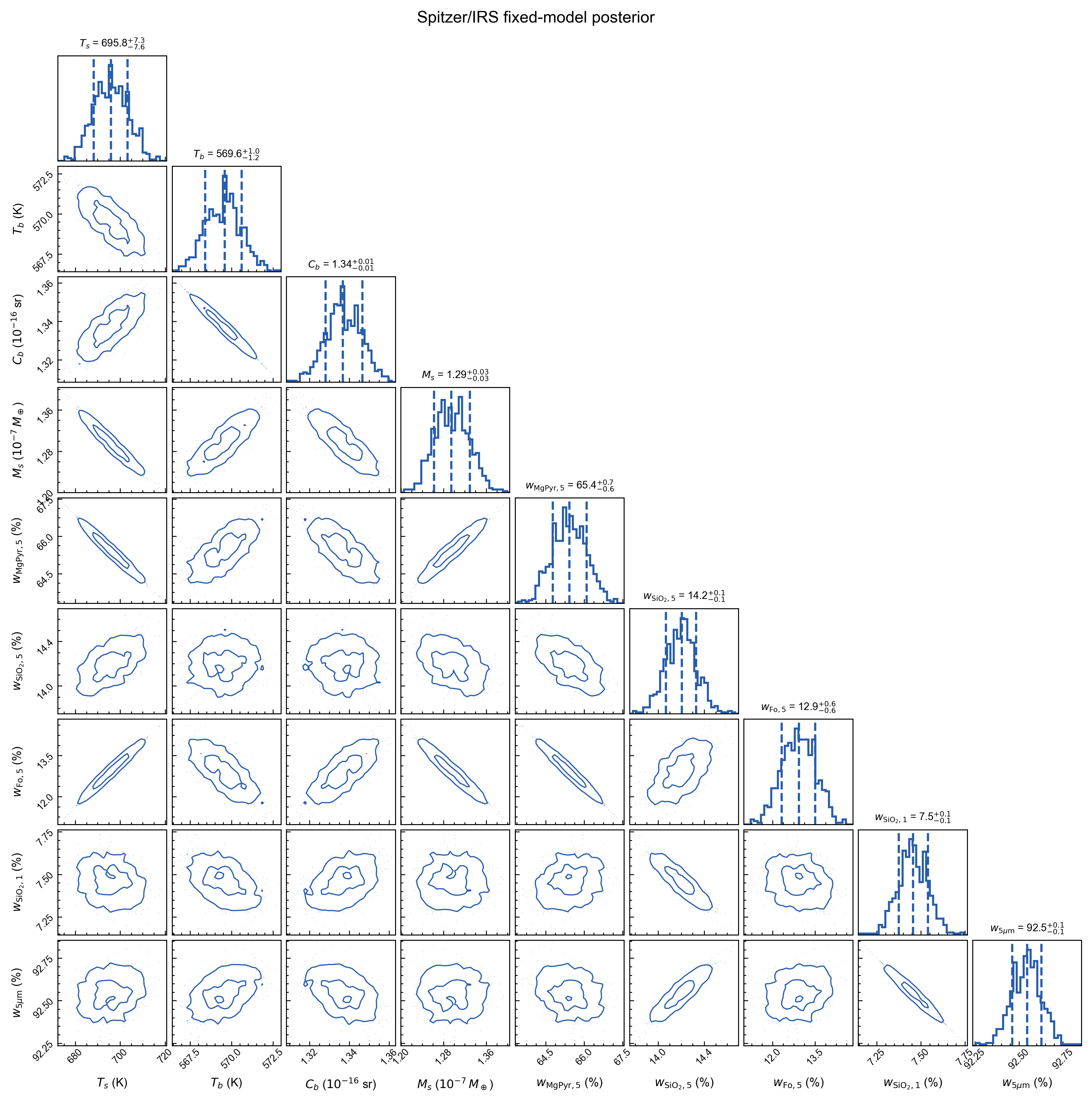}
\caption{Formal fixed-model posterior corner plot for the Spitzer/IRS fit. The variables are the quantities reported in Table~\ref{tab:free_results}: $T_s\equiv T_{\rm surf}$, $T_b\equiv T_{\rm base}$, $C_b\equiv C_{\rm base}/10^{-16}$ sr, $M_s\equiv M_{\rm surf}/10^{-7}M_\oplus$, and normalized mineral weights in percent. Here Fo denotes forsterite and $w_{5\mu{\rm m}}$ is the summed 5.0~$\mu$m weight. The one-dimensional panels mark the 16th, 50th, and 84th percentiles; the two-dimensional panels show the corresponding post-fit sample correlations. These distributions are conditional on the adopted no-rim two-isothermal model, fixed GRF opacity basis, fixed grain-size grid, and adopted base/surface continuum decomposition, and do not include opacity-library, grain-size-grid, model-selection, continuum-decomposition, or absolute-flux-calibration systematics.}
\label{fig:table1_posterior_spitzer}
\end{figure*}

\begin{figure*}[t!]
\centering
\includegraphics[width=0.95\textwidth,height=0.84\textheight,keepaspectratio]{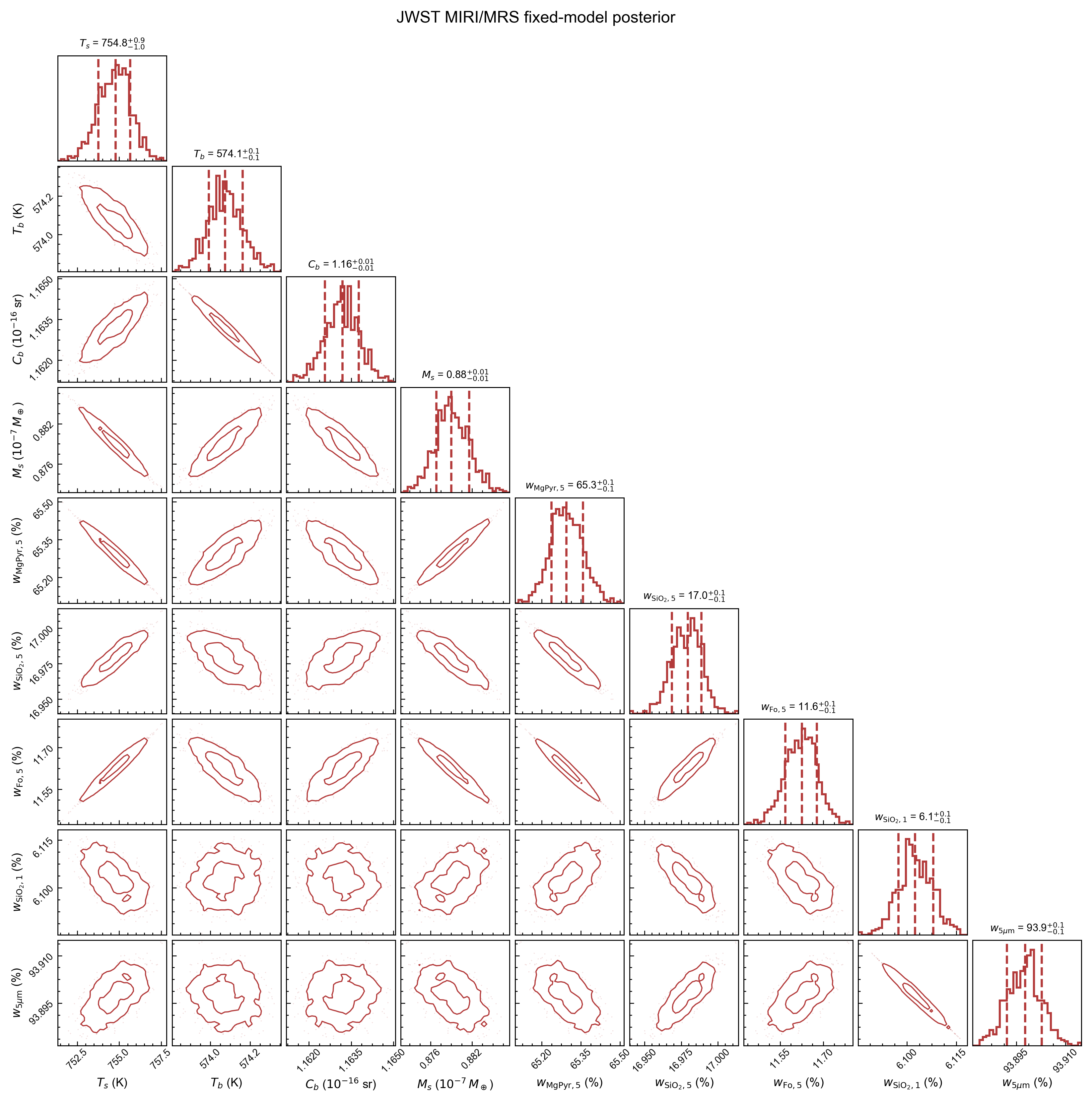}
\caption{Same as Figure~\ref{fig:table1_posterior_spitzer}, but for the JWST MIRI/MRS fit. The mass axis uses $M_s\equiv M_{\rm surf}/10^{-7}M_\oplus$, as labeled.}
\label{fig:table1_posterior_jwst}
\end{figure*}

Figures~\ref{fig:table1_posterior_spitzer} and \ref{fig:table1_posterior_jwst} show the formal fixed-model posterior distributions for the Spitzer/IRS and JWST MIRI/MRS fits, respectively.

\clearpage

\begin{acknowledgments}
We thank the referee for constructive comments that helped improve the manuscript.
This work was supported by the National Natural Science Foundation of China (NSFC) through the projects 12373028, 12322306, 12133002, and 12173047. S. W. and X. C. acknowledge support from the Youth Innovation Promotion Association of the CAS (grant Nos. 2023065 and 2022055). This work is based in part on observations made with the NASA/ESA/CSA James Webb Space Telescope (JWST). The data were obtained from the Mikulski Archive for Space Telescopes (MAST) at the Space Telescope Science Institute (STScI), which is operated by the Association of Universities for Research in Astronomy, Inc., under NASA contract NAS 5-03127 for JWST. These observations are associated with program 1206 and are available as \dataset[JWST MIRI/MRS observations of HD 15407A]{\doi{10.17909/5jkj-gx74}}.
This work is also based in part on archival data obtained with the Spitzer Space Telescope, operated by the Jet Propulsion Laboratory (JPL), California Institute of Technology under a contract with NASA, with support provided by an award issued by JPL/Caltech. Furthermore, this publication makes use of data products from the Wide-field Infrared Survey Explorer (WISE), including the AllWISE Data Release; WISE is a joint project of the University of California, Los Angeles, and JPL/Caltech, funded by NASA.
This work has made use of data from the European Space Agency mission \textit{Gaia}, processed by the \textit{Gaia} Data Processing and Analysis Consortium. This publication also makes use of data products from the Two Micron All Sky Survey, a joint project of the University of Massachusetts and the Infrared Processing and Analysis Center/California Institute of Technology, funded by NASA and the National Science Foundation.
\end{acknowledgments}

\facilities{IRAS, AKARI (IRC), Spitzer (IRS), JWST (MIRI), WISE, Gaia, 2MASS}

\software{DuCK/DuCKLinG \citep{Kaeufer2024}}

\bibliography{sample701}{}
\bibliographystyle{aasjournalv7}

\end{document}